# What are Physical States?[†]

Stephen Boughn[*]

Department of Physics, Princeton University, Princeton NJ
Departments of Physics and Astronomy, Haverford College, Haverford PA

**Abstract**

The concept of the *physical state* of a system is ubiquitous in physics but is usually presented in terms of specific cases. For example, the state of a point particle of mass *m* is completely characterized by its position and momentum. There is a tendency to consider such states as "real", i.e., as physical properties of a system. This rarely causes problems in classical physics but the notion of *real quantum states* has contributed mightily to the philosophical conundrums associated with quantum mechanics. The Einstein-Rosen-Podolsky paradox is a prime example. In fact, quantum states are not physical properties of a system but rather subjective descriptions that depend on the information available to a particular observer. This realization goes a long way toward resolving such dilemmas as Schrödinger's cat, wave function collapse, quantum non-locality, and parallel universes.

**Introduction**

The notion of the physical state of a system is a crucial concept in physics as well as in other disciplines. For macroscopic objects, quantities like temperature, pressure, composition, etc., are sufficient to characterize a state. Of course, such descriptions leave out a great deal of information about a system, in particular, the details of the microscopic constituents of the system. However, if a system is comprised of a single or of a few particles, both classical and quantum mechanics assume that a proper physical state includes all possible information about the system. In the case of a quantum system, Einstein famously concluded that the quantum state (quantum wave function) "does not provide a complete description of the physical reality". Bohr, and probably many other physicists, would disagree. The following is my pragmatic take on the completeness of such physical states.

---

[†] This essay should be considered a companion to "Against 'Reality' in Physics" (Boughn 2019)
[*] sboughn@haverford.edu

**Physical States as Complete Descriptions**

For a specific observer, the *physical state* of a system, whether it be a classical or a quantum system, is the *complete* description of the system consistent with the information about the system possessed by the observer. By "complete" I mean that from such a description, the observer can predict the results of all possible observations of the system, whether they be precise or statistical in nature. Furthermore, the dynamical theory of the system can be used to compute the time evolution of the physical state and, therefore, to predict the results of all possible future observations. Note that this definition of physical state is specific to a particular observer. A different observer with different knowledge of the system might well assign a different physical state to the system and that state can also be considered complete provided that it can be used to predict the results of all possible observations of the system *by that observer*. Therefore, the physical state is not a physical property of a system but rather is an observer dependent yet complete description of that system.

In classical mechanics, one might insist that there is a *true* physical state of a system but some observers might simply lack the information necessary to identify the true state and, therefore, can only specify an approximate state based on the information available to them. Classical mechanics, even statistical mechanics, is amenable to the notion of hidden variables, and so permits this interpretation. On the other hand, such a declaration of "truth" is metaphysical in the sense that it adds nothing to the utility of the concept of physical state delineated in the previous paragraph. If the specified state enables the observer to make accurate predictions (whether exact or statistical) of all future measurements, then that state satisfies the criterion of being a complete description. On the other hand, an "approximate state" might imply predicted results that are inconsistent with what is actually observed. For example, if the approximate state indicates a given uncertainty in position, e.g., a Gaussian distribution, but the observations of such states result in a precise value of the position or a distribution of positions that is narrower than that implied by the approximate state, then the specified state is simply not the complete description required by a "physical state" of the system (as defined above).

In quantum mechanics, the notion of a *true* physical state of a system runs into trouble. The quintessential example is an entangled state. If two components of an entangled system are spatially separated, then an observation of one of the subsystems results in new information for that observer, in which case the observer is entitled to update the quantum state describing the entangled pair. This is euphemistically referred to as "state reduction" or "collapse of the wave function". If there were, indeed, a true physical state, one is tempted to consider state reduction as a physical process, in which case that part of the state describing the remote entangled subsystem will also be instantaneously reduced, a seeming violation of relativity. This is the source for claims of quantum non-locality. On the other hand, as long as the remote observer lacks knowledge of the results of the observed subsystem, the original quantum state will provide a *complete* description of all future observations, *by that observer*, of the remote subsystem. In other words, the notions of a true physical state and state reduction are again metaphysical appendages that add nothing to the utility of the state concept.

**The EPR Paradox**

The 1935 paper by Einstein, Podolsky, and Rosen (EPR) considered the case of an entangled pair of particles. The states of the two particles were presented not as descriptions by observers but rather as "descriptions of reality", i.e., *true* physical states. EPR then find a contradiction in these assumptions and conclude: "the description of reality as given by a wave function is not complete."

Let's investigate the entangled state scenario in more detail using the Bohm spin singlet system, the same system employed by John Bell in his seminal paper on the EPR paradox (Bell 1964). The system consists of the emission of two oppositely moving spin ½ particles in a singlet state, that is, the spins of the two particles are precisely opposite each other but in an undetermined direction. According to quantum mechanics, their combined wave function (quantum state) is given by

$$\Psi(1,2) = \tfrac{1}{\sqrt{2}}\{|1,\uparrow\rangle_z|2,\downarrow\rangle_z - |1,\downarrow\rangle_z|2,\uparrow\rangle_z\} \qquad (1)$$

where ↑ and ↓ indicate the up and down *z* components of the spins of particles 1 and 2. Now suppose that the spin of particle 1 is measured with a Stern-Gerlach apparatus oriented in the $\hat{z}$ direction and is determined to be ↑. Then we "know" that the *z* component of the spin of particle 2 will be ↓ and the "true" state of particle 2 will be instantaneously reduced to $\Psi(2) = |2,↓\rangle_z$. On the other hand, the original wave function can also be expressed in terms of the *x* components (or in any direction for that matter) of the spin,

$$\Psi(1,2) = \frac{1}{\sqrt{2}}\{|1,↑\rangle_x|2,↓\rangle_x - |1,↓\rangle_x|2,↑\rangle_x\}.$$

Then, if the spin of particle 1 is measured with a Stern-Gerlach apparatus oriented in the $\hat{x}$ direction and is determined to be ↑, the spin of particle 2 will be ↓, and the "true" state of particle 2 becomes $\Psi(2) = |2,↓\rangle_x$, a state fundamentally distinct from $|2,↓\rangle_z$. The immediate transition of particle 2 to either the state $|2,↓\rangle_z$ or $|2,↓\rangle_x$, depending on the measurement performed on the distant particle 1, is the basis for claims of "action at a distance". Both Einstein and Bell accepted this picture but while Einstein was abhorred by such "spooky action at a distance", Bell embraced it as a demonstration of the "gross non-locality of nature" (Bell 1975).

Now let's look at this scenario from the perspective of a subjective description of "state". For the entangled singlet state considered above, Einstein would conclude particle 2 has no unique (real) state. Let me push back on this conclusion. After their emission, the polarization of neither particle is known. Therefore, the two particles can be considered as two unpolarized particle beams. For the observer of particle 2, this can be represented by a 50% mixture of spin up, $|2,↑\rangle$, and spin down, $|2,↓\rangle$, states (in any direction), a so-called *mixed state*. A mixed state cannot be described by a pure quantum state (e.g., Schrödinger wave function or vector in Hilbert space) but is well described by its associated density matrix $\varrho$,

$$\rho(2) = \frac{1}{2}|2,↑\rangle\langle 2,↑| + \frac{1}{2}|2,↓\rangle\langle 2,↓| \qquad (2)$$

with a similar expression for particle 1. This density matrix completely characterizes the spin measurements of particle 2 with no mention of what measurements are made on the other particle. For example, Eq. (2) implies that a measurement of the spin component in any direction will yield ↑ for 50% of the measurements and ↓ for 50% of the measurements subject to the usual statistical fluctuations. The measurement made on particle 2 is completely independent of any measurement carried out on particle 1.[1]

Okay, so far so good. Nevertheless, you may object that these single particle mixed states are mute on the correlations of measurements made on the two states. Fair enough, but now you are asking a different question. That is, after making these independent measurements, what are the correlations between them? To answer this question, to be sure, one needs the pure entangled state of Eq. 1 but this expression in no way implies that the measurements made on particle 1 have any effect whatsoever on the measurements made on particle 2 or that individually these measurements are other than those predicted by the density matrix of Eq. 2.

So after the measurement of particle 1, what is the quantum state of particle 2? Is it still the density matrix $\rho(2)$ of Eq. 2 or is it the wave function $|2,\downarrow\rangle_z$? The premise of this question is wrong. The quantum state is not an "it" in this sense. The only "it" is the quantum system, the spin ½ particle. The quantum state is not a physical property but rather a description of a physical object, and that description depends on the information available to a particular observer. For Observer 1 who has just observed particle 1 to be in the ↑ state, the available information leads to $\Psi(2) = |2,\downarrow\rangle_z$ while for Observer 2 the appropriate state is still given by the density matrix $\rho(2)$. For a probabilistic theory like quantum mechanics, confirmation can only come through the measurements of an ensemble of identically prepared systems. For observer 2, the results of such measurements will be completely described by the density matrix of Eq. 2. Precisely the same is the case for Observer 1 prior to the observation of particle 1. After the measurement, Observer 1 is justified in specifying the state of particle 2 as either $|2,\uparrow\rangle$ or $|2,\downarrow\rangle$, depending on the result that measurement. Finally, suppose that the results obtained by Observer 1 are communicated (subluminally, of course) to Observer 2. With this additional information in hand, Observer 2 will naturally specify $\Psi(2) = |2,\downarrow\rangle$ or $|2,\uparrow\rangle$ to be the quantum state of particle 2 depending on this information. The point is, a quantum state is a subjective description that depends on the information available to a specific observer.

---

[1] After all, the measurement made on particle 2 could be made in advance of even selecting the type of measurement to be made on particle 1.

**Entangled Classical State**

There is a simple classical analog of an entangled two-particle state. Suppose there are two balls, a white ball and a black ball. The two are randomly placed in either of two boxes. The boxes are closed and sent off in opposite directions where they encounter observers who open the boxes. In this case, the entangled state is a 50/50 mixture of a white ball in box 1 – black ball in box 2 and a black ball in box 1 – white ball in box 2. On the other hand, it's clear that each observer has a 50/50 chance of finding a white ball and a 50/50 chance of finding a black ball and so is justified in describing their subsystem as a 50/50 mixtures of white ball in the box and black ball in the box. This can be confirmed by compiling the data from multiple (random) trials of the experiment. There is no set of observations performed on an ensemble of random trials that disagrees with this prediction, i.e., the 50/50 mixture of white and black balls constitutes a complete description of boxes 1 and 2. If Observer 1 finds a white ball, that observer is justified in "collapsing" the entangled state to a white ball in box 1 and a black ball in box 2. However, until this information is transmitted to Observer 2, the complete description of that observer is still give by a 50/50 mixture of white and black balls in the box and this describes the predictions of future observations of box 2.

As in Bohm's spin singlet quantum system, it is also clear that the findings of the two observers will be completely correlated; if the observer of box 1 finds a white ball, the observer of box 2 will necessarily find a black ball and vice versa. So for an observer privy to the results of the observations of both boxes, the appropriate description is the mixed entangled state. As in the quantum case, we would never claim that the act of observing a white ball in box 1 *causes* a black ball to appear in box 2. Einstein would not be bothered by this situation because, in his view, there is a *true* physical state of each subsystem and so there is no necessity for positing the physical process of state reduction. We've already noted that such an assertion is completely compatible with classical physics. However, as was pointed out above, these considerations are metaphysical and don't add to the utility of the notion of a physical state.[2] (Of course, quantum

---

[2] This toy model of an entangled classical state may seem a bit contrived; however, there is a well-defined probabilistic formulation of classical mechanics based on the Hamilton-Jacobi formalism on configuration space wherein the notion of entangled states and mixtures is completely natural. (Hall & Reginatto 2016)

entanglement is a much richer phenomenon than classical entanglement and exhibits all the aspects of quantum interference with which we are familiar.)

**Final Remarks**

The notion of a physical state as a "complete description" need not encompass all the information about the system. For example, one might characterize the position and momentum of an electron with a spatial wave function that neglects any mention of the spin state of the electron. In this case, the wave function can still be considered to provide a complete description so long as observations are limited to position and momentum and there are no magnetic fields that might couple spin with the spatial trajectory. Similarly, the white ball/black ball classically entangled state presented above can still be considered complete so long as the plethora of other properties of the balls (e.g., mass, density, temperature, composition, etc.) are irrelevant to subsequent observations of the system.

In the abstract I indicated that characterizing physical states as subjective descriptions "goes a long way toward resolving such dilemmas as Schrödinger's cat, wave function collapse, quantum non-locality, and parallel universes"[3]. Wave function collapse and quantum non-locality were addressed directly in the paper. It should be clear how to apply the notion of subjective states in the resolution of other conundrums in physics, and particularly in quantum mechanics, but I'll leave this task to the reader.

**Acknowledgement**

Thanks to Marcel Reginatto who introduced me to the wonders of ensembles on configuration space and the associated probabilistic formulation of classical mechanics.

---

[3] I've discussed Hugh Everett's "many worlds interpretation" in detail elsewhere. (Boughn 2018)